\documentclass[10pt,aps,prl,floatfix,twocolumn,nofootinbib,superscriptaddress,preprintnumbers]{revtex4-2}

\usepackage{graphicx}
\usepackage{comment}
\usepackage{dcolumn}
\usepackage{bm}
\usepackage{physics}
\usepackage{amsmath,amssymb,xcolor,float,mathrsfs} 
\usepackage{xparse,etoolbox}
\usepackage{etaremune}
\usepackage[colorlinks,allcolors=blue]{hyperref} 
\usepackage[capitalize]{cleveref}
\usepackage[normalem]{ulem}

\newcommand{\submitletter}{}

\newcommand{\compile}{}

\newcommand{\hatn}{\hat{\bm{n}}}
\newcommand{\iu}{{i\mkern1mu}} 
\newcommand{\rmd}[2]{\,{\rm d}^{#1} #2 \,} 
\newcommand{\Int}[3]{%
	\ifstrempty{#3}%
	{\int \!\! \rmd{#1}{#2} \,}%
	{\int \!\! \frac{\rmd{#1}{#2}}{#3} \,}%
}%
\newcommand{\mass}{m_{\phi}}

\usepackage{Macro_TN} 

\begin{document}

\ifdefined\compile

\preprint{RESCEU-11/24}
\preprint{TU-1232}

\title{\texorpdfstring{$n\pi$}{pi} phase ambiguity of cosmic birefringence}

\author{Fumihiro Naokawa}
\affiliation{Research Center for the Early Universe, The University of Tokyo, Bunkyo-ku, Tokyo 113-0033, Japan}
\affiliation{Department of Physics, Graduate School of Science, The University of Tokyo, Bunkyo-ku, Tokyo 113-0033, Japan}
\author{Toshiya Namikawa}%
\affiliation{Center for Data-Driven Discovery, Kavli IPMU (WPI), UTIAS, The University of Tokyo, Kashiwa 277-8583, Japan}%
\author{Kai Murai}%
\affiliation{Department of Physics, Tohoku University, Sendai, Miyagi 980-8578, Japan}%
\author{Ippei Obata}%
\affiliation{Kavli IPMU (WPI), UTIAS, The University of Tokyo, Kashiwa 277-8583, Japan}%
\author{Kohei Kamada}%
\affiliation{School of Fundamental Physics and Mathematical Sciences, Hangzhou Institute for Advanced Study, University of Chinese Academy of Sciences (HIAS-UCAS), Hangzhou 310024, China}
\affiliation{International Centre for Theoretical Physics Asia-Pacific (ICTP-AP), Beijing/Hangzhou, China}
\affiliation{Research Center for the Early Universe, The University of Tokyo, Bunkyo-ku, Tokyo 113-0033, Japan}


\begin{abstract}
We point out that the rotation angle $\beta$ of cosmic birefringence, which is a recently reported parity-violating signal in the cosmic microwave background (CMB), has a phase ambiguity of $n\pi \,(n\in\mathbb{Z})$. This ambiguity 
has a significant impact on the interpretation of the origin of cosmic birefringence.
Assuming an axion-like particle as the origin of cosmic birefringence, this ambiguity can be partly broken by the anisotropic cosmic birefringence and the shape of the CMB angular power spectra. We also discuss constraints on $\beta$ from existing experimental results.
$~~~~~~~~~~~~~~~~$
\end{abstract}

\maketitle

\fi

\ifdefined \submitletter
  {\it Introduction.---}
\else
  \section{Introduction} \label{sec:intro}
\fi
Recent analyses of cosmic microwave background (CMB) polarization data have found a tantalizing hint of cosmic birefringence \cite{Komatsu:2022nvu}---a phenomenon where light traveling through space experiences a rotation of its polarization plane---with a statistical significance of $3.6\,\sigma$ \cite{Minami:2020odp,Diego-Palazuelos:2022dsq,Eskilt:2022wav,Eskilt:2022cff,Eskilt:2023ndm}. 
This parity-violating signature is a smoking gun of new physics beyond the $\Lambda$ Cold Dark Matter Model and the Standard Model of particle physics~\cite{Nakai:2023zdr}.  

Cosmic birefringence can be attributed to the presence of a pseudoscalar field of ``axion-like'' particle (ALP). 
In general, an ALP can have the Chern-Simons coupling with the photon as $\mathcal{L}\supset-g\phi F_{\mu\nu}\tilde{F}^{\mu\nu}/4$, where $g$ is a coupling constant, $\phi$ is the axion-like field, $F_{\mu \nu}$ is the electromagnetic tensor, and $\tilde{F}^{\mu \nu}$ is its dual.
Through this coupling, the ALP can cause the rotation of the polarization plane. In particular, the time evolution of the homogeneous axion-like field, $\bar{\phi}(t)$, induces isotropic cosmic birefringence.
The total rotation angle is given by 
\begin{equation}
    \label{24031403}
    \beta
    =\frac{g}{2}\int^{t_0}_{t_*} \mathrm{d}t \, \frac{\mathrm{d}\bar{\phi}(t)}{\mathrm{d}t}
     = \frac{g}{2}\left[\bar{\phi}(t_0)-\bar{\phi}(t_*)\right]\equiv \frac{g}{2} \Delta \bar{\phi},
\end{equation}
where $t_0$ is the present time, and $t_*$ is the last scattering time of CMB photons. Therefore, measuring $\beta$ with high precision results in precisely determining the model parameters and dynamics of the ALP \cite{Fujita:2020ecn,Takahashi:2020tqv,Fung:2021wbz,Nakagawa:2021nme,Jain:2021shf,Choi:2021aze,Obata:2021nql,Nakatsuka:2022epj,Lin:2022niw,Gasparotto:2022uqo,Lee:2022udm,Jain:2022jrp,Murai:2022zur,Gonzalez:2022mcx,Qiu:2023los,Eskilt:2023nxm,Namikawa:2023zux,Gasparotto:2023psh,Ferreira:2023jbu}.

In the cosmic birefringence analyses, we have estimated $\beta$ based on the relationship between the CMB angular power spectra in the presence of cosmic birefringence and those without cosmic birefringence as we describe below~(see also \cite{Lue:1998mq,Feng:2006dp}).
To derive the relationship, we first express an observed CMB polarization in the presence of cosmic birefringence as
\begin{equation}
    P^o(\hatn) = e^{2\iu\beta}P(\hatn) 
    \,, \label{Eq:QU-rot}
\end{equation}
where $P\equiv Q+iU$ with $Q$ and $U$ denoting the Stokes $Q$ and $U$ parameters of linear polarization, respectively, $\hatn$ is a line-of-sight unit vector, and $\beta$ is a rotation angle of the polarization plane. 
The superscript $o$ denotes the observed quantities at present, and $P$ on the right-hand side is the intrinsic polarization. 
In CMB analysis, we decompose CMB polarization into the parity-even $E$-modes and parity-odd $B$-modes as \cite{Zaldarriaga:1996xe, Kamionkowski:1996ks}
\begin{equation}
    E_{lm}\pm\iu B_{lm} = - \Int{2}{\hatn}{} _{\pm 2}Y^*_{lm}(\hatn) [Q\pm\iu U] (\hatn)
    \,, 
    \label{Eq:EB-def}
\end{equation}
where $_{\pm 2}Y_{lm}(\hatn)$  is the spin-2 spherical harmonics.
In the presence of isotropic cosmic birefringence, they are modified as
\begin{align}
    E^o_{lm} &= E_{lm}\cos(2\beta) - B_{lm}\sin(2\beta)
    \,, \\ 
    B^o_{lm} &= E_{lm}\sin(2\beta) + B_{lm}\cos(2\beta)
    \,, 
\end{align}
while the coefficient of a spherical harmonics expansion of temperature fluctuations,$~\Theta_{lm}= \Int{2}{\hatn}{} Y^*_{lm}(\hatn)\Theta(\hatn)~$, is not modified by the rotation.
The observed angular power spectra relevant to cosmic birefringence analysis are given by
\begin{align}
    \label{24031401}
    C_l^{\Theta E,o} &= C_l^{\Theta E}\cos(2\beta)
    \,, \\
    \label{24031402}
    C_l^{\Theta B,o} &= C_l^{\Theta E}\sin(2\beta)
    \,, \\
    \label{24011906}
    C_l^{EE,o} &= C_l^{EE}\cos^2(2\beta) + C_l^{BB}\sin^2(2\beta)
    \,, \\
    \label{24011907}
    C_l^{BB,o} &= C_l^{EE}\sin^2(2\beta) + C_l^{BB}\cos^2(2\beta)
    \,, \\
    \label{24011908}
    C_l^{EB,o} &= \frac{1}{2}(C_l^{EE}-C_l^{BB})\sin(4\beta)
    \,. 
\end{align}
 Here we have assumed no intrinsic $EB$ and $\Theta B$ correlations. From the last three equations, we obtain the relationship between $\beta$ and the observed CMB polarization spectra as \cite{Zhao:2015mqa}
\begin{equation}
    \label{23041102}
    \tan(4\beta)=
    \frac{2C_l^{EB,o}}{C_l^{EE,o}-C_l^{BB,o}}.
\end{equation}
Therefore, we can estimate $\beta$ from the above equation, and previous works have reported $\beta \sim 0.3$~deg \cite{Minami:2020odp,Diego-Palazuelos:2022dsq,Eskilt:2022wav,Eskilt:2022cff,Eskilt:2023ndm}. 

One important effect that has been overlooked in previous work is the possibility of multiple rotations. All observed power spectra in Eqs.~\eqref{24031401}--\eqref{24011908} are invariant under $\beta \to \beta \pm \pi$, and the general solution of $\beta$ is given by 
\begin{equation}
\label{24011910}
    \beta = \frac{1}{4}\arctan
    \left(
    \frac{2C_l^{EB,o}}{C_l^{EE,o}-C_l^{BB,o}}
    \right)
    +n\pi~~~(n\in\mathbb{Z}).
\end{equation}
The integer parameter $n$ is a convenient index for the number of rotations. This $n\pi$ phase ambiguity\footnote{This kind of ambiguity has been a problem 
in the context of Faraday rotation to reveal cosmic magnetic field and is called ``$n\pi$-ambiguity''~(e.g.~\cite{Van:Eck_2021,1975AA43233V}). As for cosmic birefringence, this ambiguity is also mentioned in Ref.~\cite{Nilsson:2023sxz}.} reflects that Stokes parameters are spin-2 quantities. Eq.~\eqref{24011910} is valid regardless of the origin of cosmic birefringence as long as all CMB photons experience a uniform birefringence angle $\beta$. 

In this work, we show how to constrain $n$ from CMB observables. 
If the rotation angle is constant in time, we cannot constrain $n$ from the CMB angular power spectrum alone. From a phenomenological point of view, however, depending on a certain ALP model as the origin of cosmic birefringence, we can limit the allowed range of $\beta$. From Eq.~\eqref{24031403}, larger $|\beta|$ means larger $g|\Delta\bar{\phi}|$, while upper
limits of $|\Delta\bar{\phi}|$ (more precisely, the density parameter of ALP, $\Omega_\phi$) and $g$ are set~\cite{Fujita:2020ecn} by the cosmological~\cite{Hlozek:2014lca,P18:main} and astrophysical observations~\cite{Berg:2016ese}, respectively. Assuming a mass potential model $V(\phi)=m_\phi^2 \phi^2/2$ for the ALP, for example, the upper limit is given by $|\beta| \lesssim 10^7~\mathrm{deg}~(|n|\lesssim10^5)$ from the condition to explain $\beta\sim0.3$ obtained in Ref.~\cite{Fujita:2020ecn}.  
In the following, we show that CMB observations can also tightly constrain $n$.

\ifdefined \submitletter
  {\it Constraint from anisotropic cosmic birefringence.---} 
\else
  \section{Constraint from anisotropic cosmic birefringence}
  \label{24011905}
\fi
The rotation angle of cosmic birefringence has anisotropic component, $\beta(\hatn)=\beta+\alpha(\hatn)$, if the axion-like field has fluctuations: 
\al{
    \alpha(\hatn) \equiv -\frac{g\delta\phi(\chi_*\hatn,\eta_*)}{2} = -\frac{g\bar{\phi}_{\rm ini}}{2}\frac{\delta\phi(\chi_*\hatn,\eta_*)}{\bar{\phi}_{\rm ini}}
    \,, 
}
where we have split the ALP as $\phi({\bm x},\eta) = \bar{\phi}(t) + \delta \phi({\bm x},\eta)$, $\chi_*$ is the comoving distance to the last scattering surface, $\bar{\phi}_{\mathrm{ini}}=\bar{\phi}(t_*)$, and $\eta_*=\eta_0-\chi_*$ is the conformal time at the last scattering with $\eta_0$ being the present conformal time. 

As $n$ increases, $g\bar{\phi}_{\rm ini}/2$ increases, and so does the amplitude of the anisotropic birefringence. On the other hand, multiple CMB observations have placed upper bounds on the amplitude of the anisotropic cosmic birefringence. This upper limit gives a constraint on $n$. 

The fluctuations of the axion-like field inevitably acquire non-zero values from the metric perturbations, if the background field exists \cite{Caldwell:2011,Capparelli:2019:CB,Greco:2022:aniso-biref-tomography}.
For a background solution of $\bar{\phi}$ determined by $m_\phi$ and $\bar{\phi}_\mathrm{ini}$, 
we solve the evolution equations of the perturbed pseudoscalar fields normalized as $ \delta\varphi\equiv\delta\phi/\bar{\phi}_{\rm ini}$ \cite{Caldwell:2011,Capparelli:2019:CB}: 
\al{
    \delta\varphi'' &+ 2\mC{H}\delta\varphi' + \left(k^2+a^2\mass^2\right)\delta\varphi 
    \notag \\
    &= \bar{\varphi}'(3\Phi'+\Psi') - 2a^2\mass\bar{\varphi}\Psi
    \,, \label{Eq:dphi:EoM}
}
where $'$ denotes the time derivative with respect to the conformal time $\eta$, $a$ is the scale factor, $k$ is the comoving Fourier wavenumber of the perturbations, $\mC{H}=a'/a$, $\bar{\varphi}=\bar{\phi}/\bar{\phi}_{\rm ini}$, and $\Phi$ and $\Psi$ are the gravitational potential in the conformal Newtonian gauge (see Ref.~\cite{Caldwell:2011} for their definitions) and are obtained from {\tt CAMB} \cite{Lewis:1999bs}. 
The adiabatic initial condition is given by 
\al{
    \delta\varphi &= \frac{1}{2}\frac{\bar{\varphi}'}{\mC{H}}\Psi 
    \,, \\
    \delta\varphi' &= -\frac{\bar{\varphi}'}{2}\Psi - \frac{1}{2}\frac{a^2\mass\bar{\varphi}}{\mC{H}}\Psi
    \,. 
}
We define the transfer function as 
\al{
    \delta\varphi(k,\eta) \equiv T_{\delta\varphi}(k,\eta){\cal R}_{\rm ini}(k)
    \,, 
}
where $\mathcal{R}_{\rm ini}$ is the primordial comoving curvature perturbation, which determines $\Phi$ and $\Psi$. 
The angular power spectrum of the anisotropic birefringence from recombination is given by \cite{Greco:2022:aniso-biref-tomography,Greco:2024}: 
\al{
    C_L^{\alpha\alpha} = \left(\frac{g\bar{\phi}_{\rm ini}}{2}\right)^2 4\pi \Int{}{(\ln k)}{} \mC{P}_{\cal R}(k) [j_L(k\chi_*)T_{\delta\varphi}(k,\eta_*)]^2
    \,,
}
with $j_L(k\chi)$ being the $L$-th spherical Bessel function and ${\cal P}_{\cal R}(k)=A_{\rm s}(k/k_*)^{n_{\rm s}-1}$, where $A_{\rm s}$ is the primordial scalar amplitude, $n_s$ is the spectral index, and the pivot scale is chosen as $k_*=0.05$ Mpc$^{-1}$.

CMB experiments have placed upper bounds on the anisotropic cosmic birefringence from the recombination, assuming that $\delta\phi$ is approximately constant during the recombination \cite{Namikawa:2020:biref,Bianchini:SPT:2020,Bortolami:2022:PR4,Zagatti:2024:PR4}. We thus restrict the mass range to $\mass\alt 10^{-28}\,$eV.  The transfer function is obtained numerically by solving Eq.~\eqref{Eq:dphi:EoM}. We compute the angular power spectrum with the Planck PR3 best-fit values of TT,TE,EE+lowE+lensing in Table 2 of Ref.~\cite{P18:main}. 

\begin{figure}
    \centering
    \includegraphics[width=80mm]{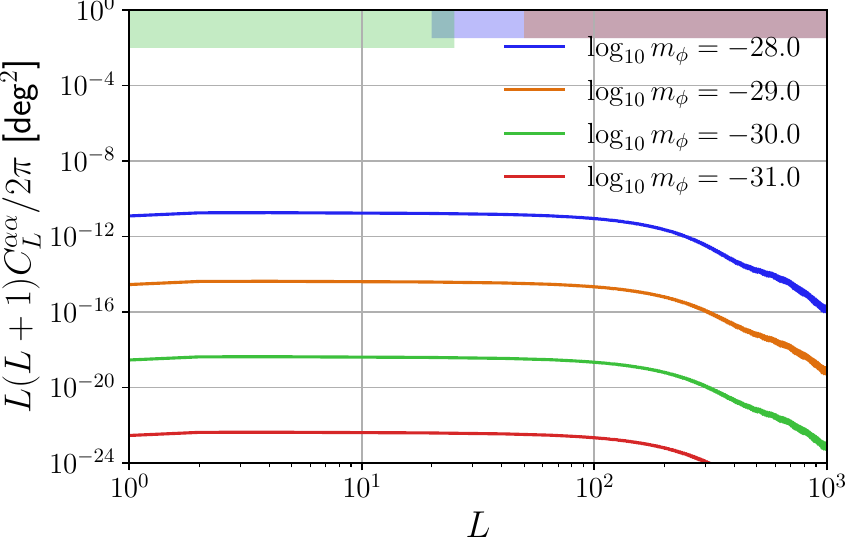}
    \caption{Angular power spectrum of anisotropic cosmic birefringence from the ALP (mass potential model) with varying $m_\phi$. We assume $g\bar{\phi}_{\rm ini}/2=0.34$\,deg and $n=0$. The shaded regions are excluded by the measurements of $C_L^{\alpha\alpha}$ by Planck (green) \cite{Bortolami:2022:PR4}, ACTPol (blue) \cite{Namikawa:2020:biref}, and SPTpol (red) \cite{Bianchini:SPT:2020}.}
    \label{fig:claa_const}
\end{figure}

Figure~\ref{fig:claa_const} shows the angular power spectrum of the anisotropic cosmic birefringence with $g\bar{\phi}_{\rm ini}/2=0.34$\,deg and $n=0$. If $n\geq 1$, the power spectrum approximately scales as $C_{L,n}^{\alpha\alpha}/C_{L,n=0}^{\alpha\alpha}\simeq  ((g\bar{\phi}_{\mathrm{ini}})_{n}/(g\bar{\phi}_{\mathrm{ini}})_{n=0})^2 \sim (180n\,{\rm deg}/0.34\,{\rm deg})^2\sim 10^5n^2$. 
We also show the latest bounds on the amplitude of the scale-invariant anisotropic birefringence power spectrum, $L(L+1)C_L^{\alpha\alpha}/2\pi\alt 10^{-3}$\,deg$^2$ at $2\,\sigma$. Using the scaling relation, and the bounds at $L\lesssim 10^2$, we find that $n\alt 10$ ($\mass=10^{-28}$\,eV), $n\alt 10^{3}$ ($\mass=10^{-29}$\,eV), $n\alt 10^{5}$ ($\mass=10^{-30}$\,eV), and $n\alt 10^{7}$ ($\mass=10^{-31}$\,eV).

\begin{figure}
    \centering
    \includegraphics[width=95mm]{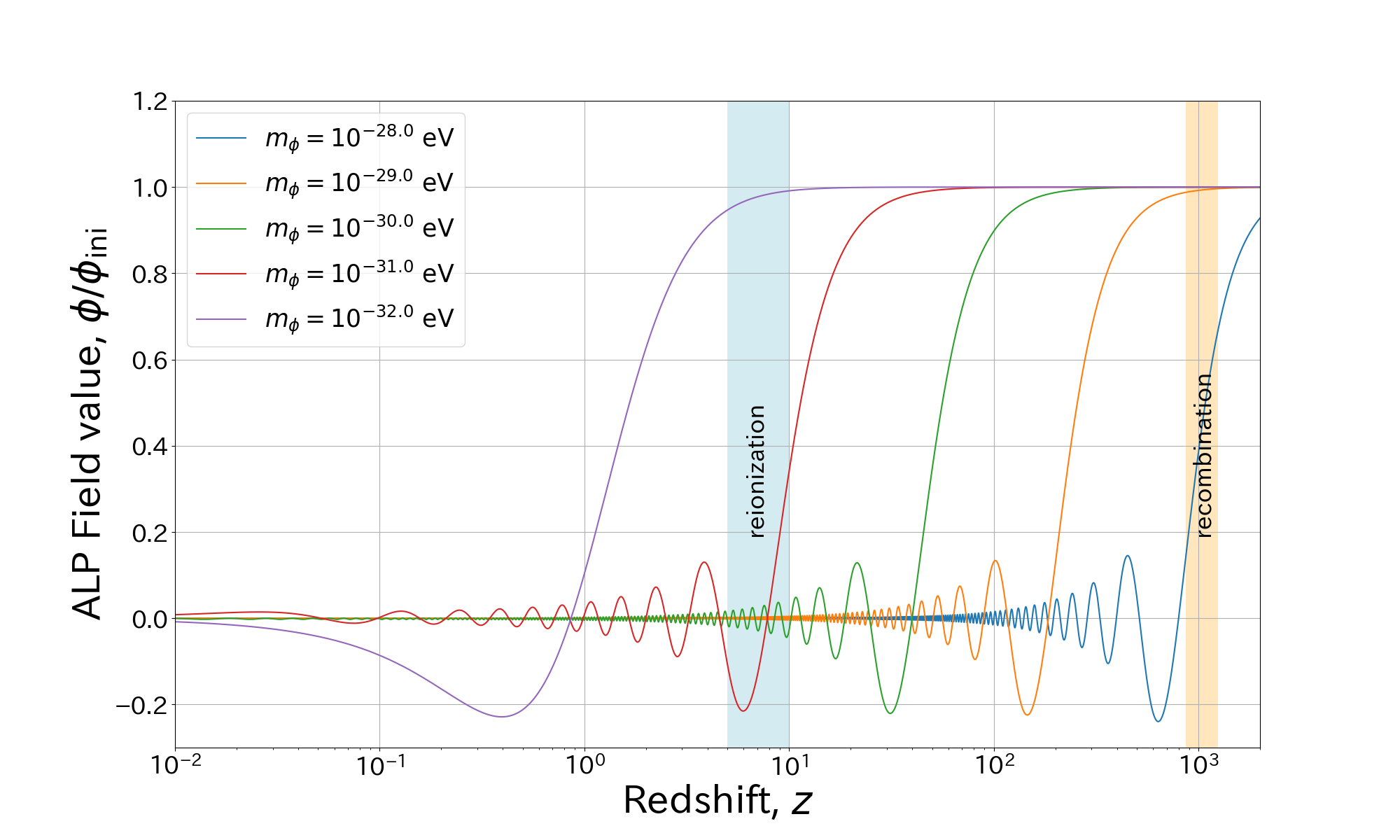}
    \caption{ALP dynamics for $10^{-32}~\mathrm{eV}<m_\phi<10^{-28}~\mathrm{eV}$. Axion-like field changes around recombination for $m_\phi\sim10^{-28}~\mathrm{eV}$ and reionization for $m_\phi\sim10^{-31}~\mathrm{eV}$}
    \label{fig:ALPdynamics}
\end{figure}

\ifdefined \submitletter
  {\it Constraint from power spectrum tomography.---}
\else
  \section{Constraint from power spectrum tomography} 
  \label{24012206}
\fi
When the constant rotation approximation is invalid, we can no longer use Eq.~\eqref{24011908} to calculate $C_l^{EB}$. 
In that case, we have to solve the Boltzmann equation of CMB polarization including the cosmic birefringence effect \cite{Liu:2006uh,Finelli:2008,Gubitosi:2014cua,Lee:2016jym}: 
\begin{align}
    \label{24012301}
    &_{\pm2}\Delta'_P + \iu q\mu {}_{\pm2}\Delta_P 
    \notag \\
    &\qquad = a n_{\rm e}\sigma_T
        \left[
            -{}_{\pm2}\Delta_P + \sqrt{\frac{6\pi}{5}}{}_{\pm2}Y_{20}(\mu)\Pi(\eta,q)
        \right]
    \notag \\
    &\qquad\qquad \pm \iu g\phi'{}_{\pm2}\Delta_P
    \,, 
\end{align}
where $n_e$ is the electron number density, $\sigma_T$ is the Thomson scattering cross section, and $\Pi$ is the polarization source term~\cite{Zaldarriaga:1996xe}.
The last term represents the effect of birefringence. The equation of motion of the homogeneous axion-like field $\bar{\phi}$ is given by 
\begin{equation}
    \bar{\phi}'' + 2\frac{a'}{a}\bar{\phi}' + a^2m_\phi^2 \bar{\phi} = 0
    \,. \label{Eq:phi-EoM}
\end{equation}
As shown in Fig.~\ref{fig:ALPdynamics}, the ALP dynamics through cosmic history depends on $m_\phi$.
After solving Eqs.~\eqref{24012301} and \eqref{Eq:phi-EoM}, we perform a spin-2 spherical harmonics expansion on $_{\pm2}\Delta_P$ as
\begin{align}
    _{\pm2}\Delta_P(\eta_0,q,\mu) 
    &\equiv -\sum_{l} \iu^{-l}\sqrt{4\pi(2l+1)} 
    \notag \\
    &\quad\times [\Delta_{E,l}(q)\pm \iu\Delta_{B,l}(q)] {}_{\pm2}Y_{l0}(\mu) 
    \,. 
\end{align}
From these coefficients, we obtain the angular power spectrum as 
\begin{equation}
    C_l^{XY} = 4\pi
    \Int{}{(\ln q)}{} \mathcal{P}_{\cal R}(q)\Delta_{X,l}(q)\Delta_{Y,l}(q)
    \,, \label{Eq:ClXY}
\end{equation}
where $X$ and $Y$ denote $E$-mode or $B$-mode.
We obtain the angular power spectra using a Boltzmann solver, \texttt{birefCLASS}~\cite{Nakatsuka:2022epj, Naokawa:2023upt, Lesgourgues:2011re}.

Reference~\cite{Nakatsuka:2022epj} revealed that the shape of $C_l^{EB}$ changes depending on the ALP dynamics. When $\bar{\phi}(t)$ substantially evolves during the recombination epoch (like $m_\phi=10^{-28}~\mathrm{eV}$ in  Fig.~\ref{fig:ALPdynamics}), the peaks of $C_l^{EB}$ at high-$l$ are shifted. This is because the birefringence angle of each photon varies depending on when it was released. 

\begin{figure}
    \centering
    \includegraphics[width=85mm]{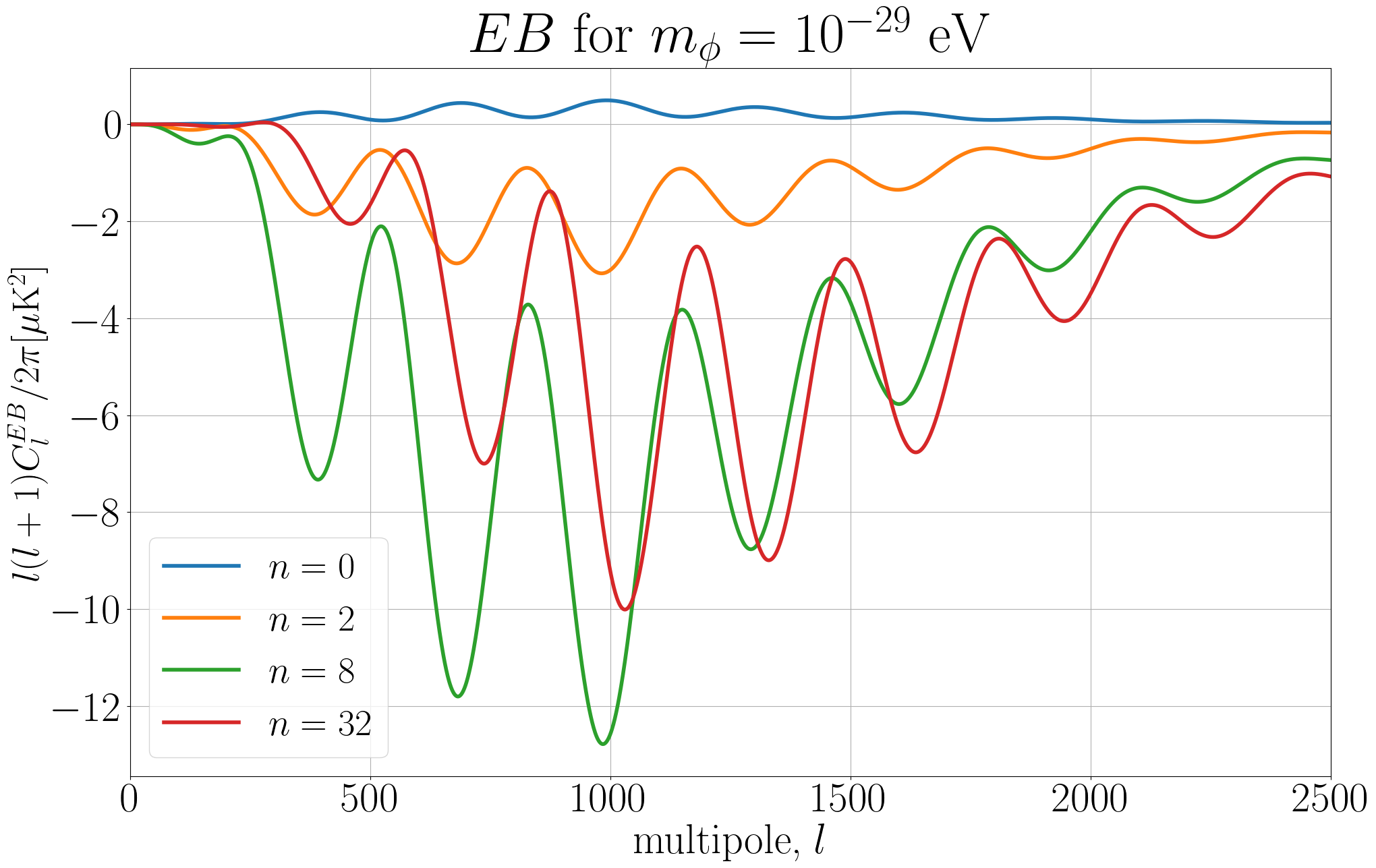}
    \caption{EB power spectrum for $m_\phi=10^{-29}~\mathrm{eV}$ for $n=0$ (blue), $n=2$ (orange), $n=8$ (green), and $n=32$ (red).}
    \label{fig:-29eV}
\end{figure}

\begin{figure}
    \centering
    \includegraphics[width=85mm]{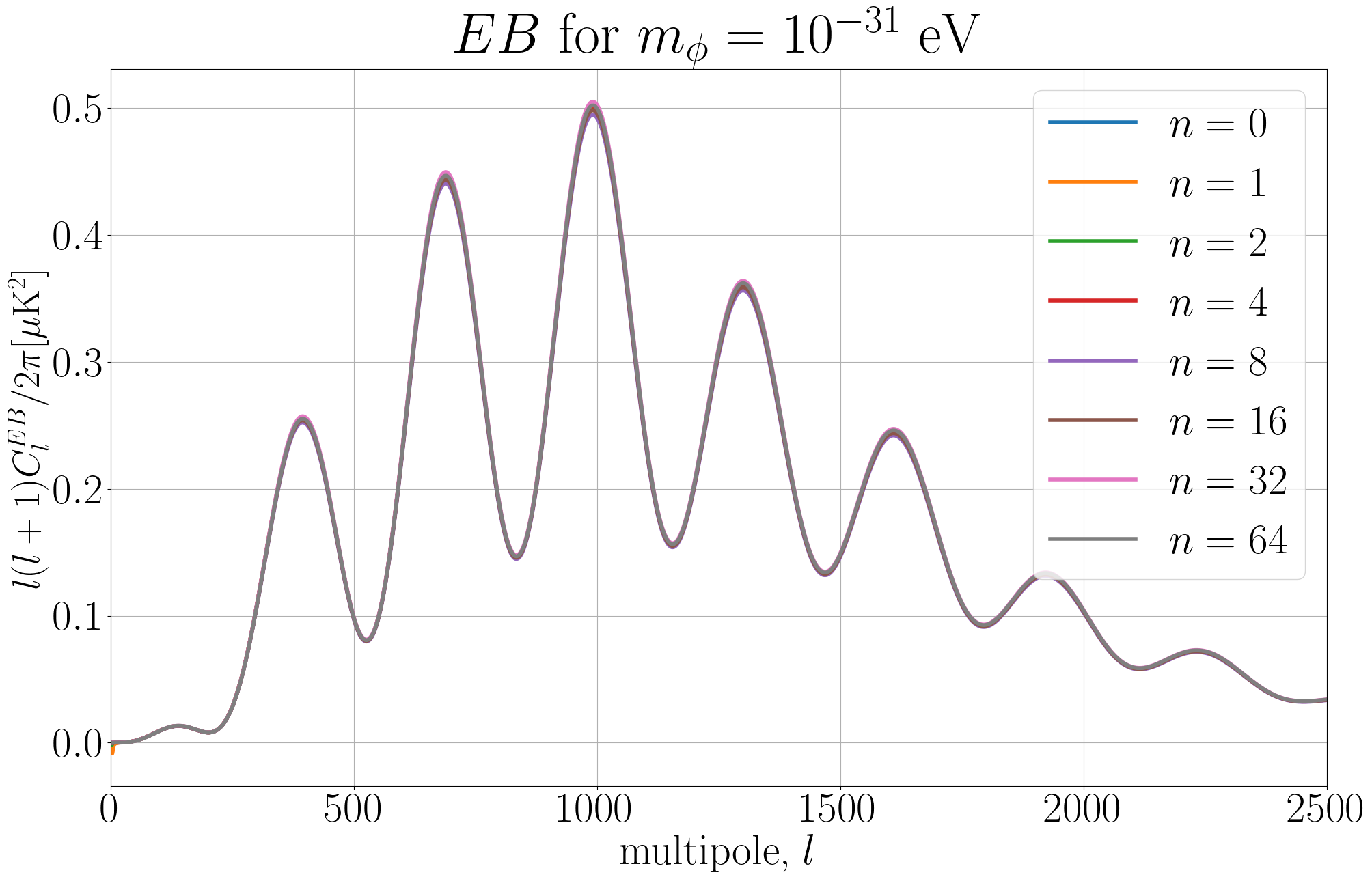}
    \caption{Same as Fig.~\ref{fig:-29eV} but for $m_\phi=10^{-31}~\mathrm{eV}$}
    \label{fig:-31eV}
\end{figure}

When $n\neq0$, this effect is amplified because $\beta$ is larger by $n\pi$ for the same amount of $\Delta\bar{\phi}$. 
In Fig.~\ref{fig:-29eV}, we show $C_l^{EB}$ for $m_\phi=10^{-29}\,\mathrm{eV}$ and various $n$. As expected, the peak positions of $C_l^{EB}$ are shifted when $n\not=0$. On the other hand, we may adjust the overall amplitude by tuning $\beta$ marginally. In the case of $m_\phi=10^{-30}\,\mathrm{eV}$, such effect is weaker, and for $m_\phi=10^{-31}\,\mathrm{eV}$ shown in Fig.~\ref{fig:-31eV}, we can no longer find any changes of $C_l^{EB}$ other than the overall amplitude.
Therefore, for $m_\phi=10^{-31}\,\mathrm{eV}$, 
all $n$ in the figure are not excluded by the shift of the peaks of $C_l^{EB}$.

\begin{figure}
    \centering
    \includegraphics[width=85mm]{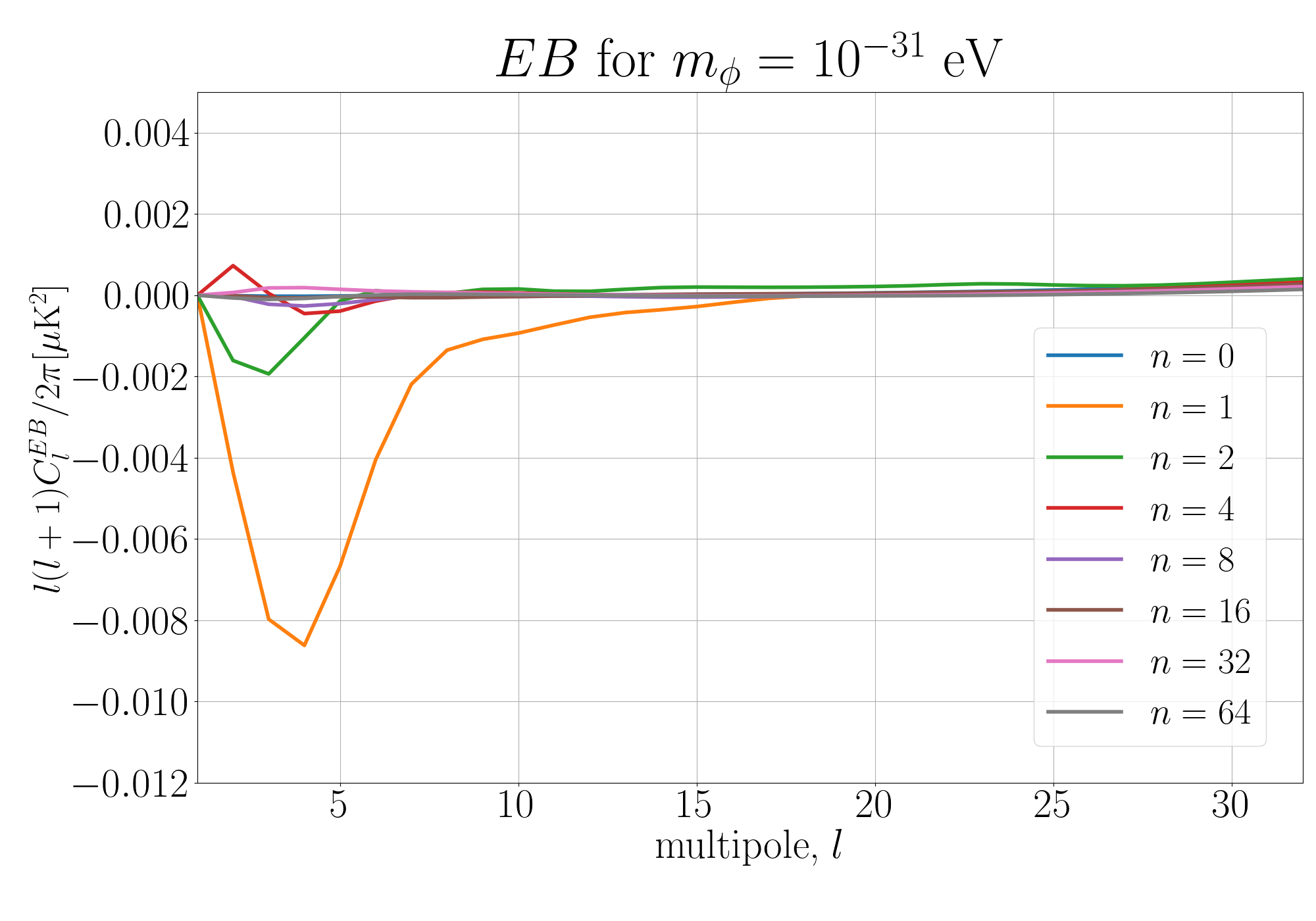}
    \caption{Low-$l$ feature of EB power spectrum for $m_\phi=10^{-31}~\mathrm{eV}$ with varying $n$. While for $n=0$, there is not a reionization bump, there are bumps for some small nonzero $n$.}
    \label{fig:-31eV_EB_lowell}
\end{figure}

However, even for such light mass $C_l^{EB}$ at low-$l$ has an interesting behavior.
Figure \ref{fig:-31eV_EB_lowell} shows the enlarged view of Fig.~\ref{fig:-31eV}, focusing on low-$l$ ($l\lesssim30$). 
The amplitudes of $C_l^{EB}$ are approximately zero for all $n$ except for $1 \leq n \lesssim 4$.
This is interpreted as follows. 
For these $n$, the oscillation of $\bar{\phi}$ during reionization is enhanced, resulting in the increased amplitude of $C_l^{EB}$ at $l \sim 3$. 
When $n\gtrsim10$, however, the corresponding birefringence angle becomes much larger than $\pi$, and $C_l^{EB}$ is washed out due to the superposition of randomly-oriented photons' polarizations emitted at the reionization epoch.

\begin{figure}
    \centering
    \includegraphics[width=85mm]{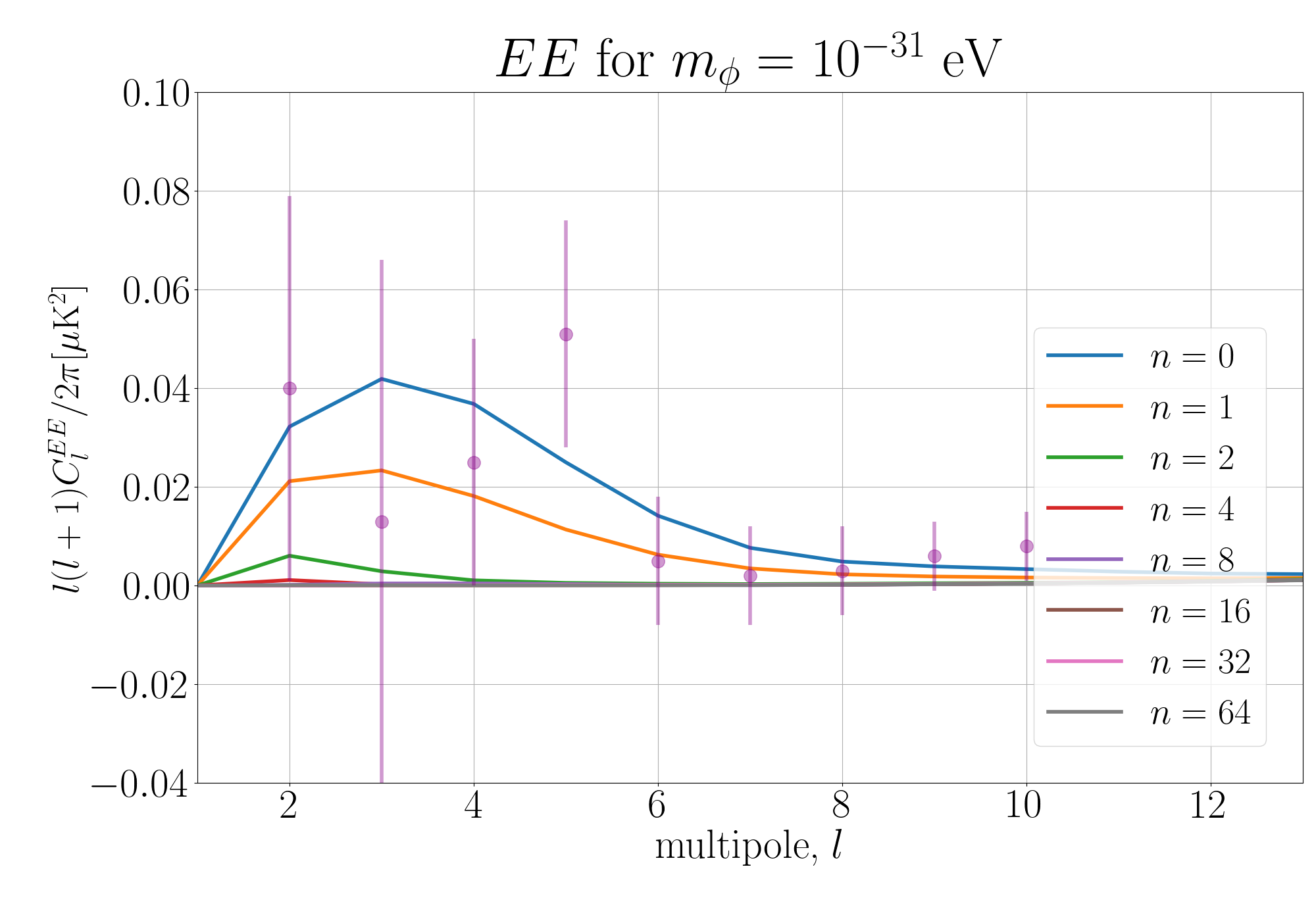}
    \caption{Low-$l$ feature of EE power spectrum for $m_\phi=10^{-31}~\mathrm{eV}$. The purple points with error bars are the latest results ($l\leq10$) by the final Planck Data Release~\cite{Tristram:2023haj}.}
    \label{fig:-31eV_EE_lowell}
\end{figure}

The washout effect also appears in $C_l^{EE}$. 
Figure~\ref{fig:-31eV_EE_lowell} shows $C_l^{EE}$ for $m_\phi=10^{-31}~\mathrm{eV}$, focusing on the reionization bump 
at $l \lesssim 10$~\cite{Zaldarriaga:1996ke} with the error bars from the Planck observation \cite{Tristram:2023haj}. 
We find that the reionization bump is suppressed with increasing $n$ and almost vanishes for $n\gtrsim10$. This behavior is also due to the washout effect by the amplified birefringence during reionization. 

$C_l^{EE}$ without the reionization bump are not excluded significantly with Planck~\cite{Tristram:2023haj}, but future full-sky observations may determine its presence or absence (also for $C_l^{EB}$).  
In other words, using the above low-$l$ features, 
we can explore the value of $n$ for relatively light ALP mass. 
Here, we only discussed the case of $m_\phi=10^{-31}~\mathrm{eV}$, but we also find 
a similar trend at low-$l$ for $m_\phi=10^{-30}~\mathrm{eV}$ and $10^{-32}~\mathrm{eV}$.

\ifdefined \submitletter
  {\it Summary and Discussion.---}
\else
  \section{Summary and Discussion}
\fi
We reexamined the 
possible rotation angle $\beta$ of cosmic birefringence including the phase ambiguity, which has been overlooked in previous studies. Due to this ambiguity, $\beta$ is not determined up to an integer parameter $n$ (Eq.~\eqref{24011910}).
Assuming that cosmic birefringence is induced by an axion-like field with a mass potential $V(\phi)= m_\phi^2 \phi^2/2$, we explored the ways to constrain $n$. We found two major constraints on $n$: one from anisotropic cosmic birefringence and the other from power spectrum tomography. The former is more sensitive to heavier ALP like $m_\phi\sim10^{-28}~\mathrm{eV}$, while the latter limits $n$ for $m_\phi \sim \mathcal{O}(10^{-31})$\,eV using the reionization bump of $C_l^{EE}$ as well as $C_l^{EB}$.
We conclude that there are a large number of overlooked solutions of $\beta$ for recent reports of cosmic birefringence, and we can test the possibility of $n \neq 0$
by the near future observations like Simons Observatory \cite{SimonsObservatory}, LiteBIRD \cite{LiteBIRD:2022:PTEP}, and CMB-S4 \cite{CMBS4:r-forecast}.

The calculation of cosmic birefringence employs an approximation, $|g_{\phi}\dot{\bar{\phi}}\lambda_{\rm CMB}|\ll 1$, where $\lambda_{\rm CMB}$ is the wave length of the CMB photon. In CMB experiments, we usually use frequency bands around $100\,$GHz corresponding to $\lambda_{\rm CMB}\sim 1\,$mm. This paper assumes ALPs whose field values change in a cosmological time scale, i.e., $g_{\phi}\dot{\bar{\phi}}\sim \Delta(g_{\phi}\bar{\phi})/\Delta\eta$ with $\Delta\eta\agt 1\,$Mpc $\sim 10^{25}\,$mm. Our analysis shows that $\Delta (g_{\phi}\bar{\phi})\alt 10^{10}$. Thus, the approximation still holds even if we allow the maximum value of $g_{\phi}\bar{\phi}_{\rm ini}$ obtained in this paper. 

We have focused on ALP-induced cosmic birefringence with the mass potential model. Similar features in the power spectra will also appear in other models of cosmic birefringence, including the early dark energy case \cite{Murai:2022zur}. 
Thus, when interpreting the results from Ref.~\cite{Eskilt:2023nxm}, we need to account for cases where $n\not=0$. 
In general, a tiny polarization rotation angle that produces negligible modifications to the power spectra is enhanced by increasing $n$, and the power spectra could be significantly modified for a nonzero $n$. 

The constraint on $n$ degenerates with the CMB optical depth $\tau$ when we use the reionization bump of $C_l^{EE}$ for $m_\phi=10^{-30}\,\text{--}\,10^{-32}$\,eV. 
A larger $\tau$ produces a larger amplitude of the reionization bump while a larger $n$ attenuates it. 
Therefore, when we use $C_l^{EE}$ to determine $n$, we have to determine $\tau$ in other ways. For example, the results of recent astronomical observations of high-z galaxies like JWST 
provide a powerful way to determine $\tau$ independently. Recent results disfavor larger $\tau$~\cite{nakane2024lyalpha, umeda2023jwst}. Performing parameter fitting, setting both $\tau$ and $n$ as free parameters, is left for future work.


\begin{acknowledgments}
We are grateful to Eiichiro Komatsu, Maresuke Shiraishi, and Masami Ouchi for fruitful discussions on this paper.  This work was supported in part by JSPS KAKENHI Grant Numbers 
JP24KJ0668 (F.\,N.), JP20H05859 and JP22K03682 (T.\,N.), JP23KJ0088 and JP24K17039 (K.\,M.), JP19K14702 (I.\,O.), and JP23K17687 (K.\,K.), and the National Natural Science Foundation of China (NSFC) under Grant No. 12347103 (K.\,K.). F.\,N. thanks the Forefront Physics and Mathematics Program to Drive Transformation (FoPM), a World-leading Innovative Graduate Study (WINGS) Program, the University of Tokyo. 
Part of this work uses resources of the National Energy Research Scientific Computing Center (NERSC). The Kavli IPMU is supported by World Premier International Research Center Initiative (WPI Initiative), MEXT, Japan. 
\end{acknowledgments}

\appendix


\bibliographystyle{apsrev4-1}
\bibliography{cite}

\end{document}